\documentclass[%
 reprint,
 amsmath,amssymb,
 aps,
]{revtex4-1}

\usepackage{graphicx}% Include figure files
\usepackage{amsmath}
\usepackage{dcolumn}% Align table columns on decimal point
\usepackage{bm}% bold math
\usepackage{color}
\usepackage[normalem]{ulem}
\usepackage{url}
\begin{document}

    \preprint{APS/123-QED}
    
    \title{Collective motion of run-and-tumble particles \\ drives aggregation in one-dimensional systems.}% Force line breaks with \\
    
    \author{C. Miguel Barriuso Gutiérrez}
    \altaffiliation{Departamento de Estructura de la Materia, F\'isica T\'ermica y Electr\'onica, Facultad de Ciencias F\'isicas, Universidad Complutense de Madrid, 28040, Madrid, Spain}
    
    \author{Christian Vanhille-Campos}
    \altaffiliation{Department of Physics and Astronomy, Institute for the Physics of Living Systems, University College London, London WC1E 6BT, United Kingdom.}
    \altaffiliation{MRC Laboratory for Molecular Cell Biology, University College London, London WC1E 6BT, United Kingdom.}
    \altaffiliation{Departamento de Estructura de la Materia, F\'isica T\'ermica y Electr\'onica, Facultad de Ciencias F\'isicas, Universidad Complutense de Madrid, 28040, Madrid, Spain}
    
    \author{Francisco Alarcón}
    \altaffiliation{Departamento de Estructura de la Materia, F\'isica T\'ermica y Electr\'onica, Facultad de Ciencias F\'isicas, Universidad Complutense de Madrid, 28040, Madrid, Spain}
    \altaffiliation{Departamento de Ingeniería Física, División de Ciencias e Ingenierías, Universidad de Guanajuato, Loma del Bosque 103, 37150 León, Mexico.}
    
    \author{Ignacio Pagonabarraga}
    \altaffiliation{Departament de F\'{\i}sica de la Mat\`eria Condensada, Universitat de Barcelona, 08028 Barcelona, Spain.}
    \altaffiliation{Universitat de Barcelona Institute of Complex Systems (UBICS), Universitat de Barcelona, 08028 Barcelona, Spain}
    \altaffiliation{CECAM, Centre Europ\'een de Calcul Atomique et Mol\'eculaire, \'Ecole Polytechnique F\'ed\'erale de Lasuanne, Batochime, Avenue Forel 2, 1015 Lausanne, Switzerland}
    
    \author{Ricardo Brito}
    \altaffiliation{Departamento de Estructura de la Materia, F\'isica T\'ermica y Electr\'onica, Facultad de Ciencias F\'isicas, Universidad Complutense de Madrid, 28040, Madrid, Spain}
    
    \author{Chantal Valeriani}%
    \altaffiliation{Departamento de Estructura de la Materia, F\'isica T\'ermica y Electr\'onica, Facultad de Ciencias F\'isicas, Universidad Complutense de Madrid, 28040, Madrid, Spain}
    \email{cvaleriani@ucm.es}
    
    
    %\affiliation{
    %\affiliation{
    %\affiliation{%
    
    
    \date{\today}% It is always \today, today,
    
    \begin{abstract}
    
     Active matter deals with systems whose particles consume energy at the individual level in order to move. 
To unravel  features such as the emergence of collective structures several models have been suggested, such as the on-lattice  model of run-and-tumble particles implemented via the Persistent Exclusion Process (PEP). 
In our work, we study a one dimensional system of run-and-tumble repulsive  or attractive particles, both on- and off-lattice. 
Additionally, we implement a cluster motility dynamics in the on-lattice case (since in the off-lattice case cluster motility arises from the individual particle dynamics). While we observe important differences between discrete and continuous dynamics, few common features are of particular importance. Increasing particle density drives aggregation across all different  systems  explored. 
For non-attractive particles, the effects of particle activity on aggregation are largely independent of the details of the dynamics.
On the contrary, once attractive interactions are introduced, the steady-state, which is completely determined by the interplay between these and the particles’ activity, becomes highly dependent on the details of the dynamics.

    \end{abstract}

    \pacs{Valid PACS appear here}                              
    % PACS, the Physics and Astronomy classification Scheme.
    % \keywords{Suggested keywords}
    % Use showkeys class option if keyword display desired

\maketitle

%%%MAIN TEXT%%%%
\section{\label{Intro}Introduction}
\noindent Active matter encompasses any kind of soft matter system where particles consume energy at the individual level to achieve motion, thus leading the system to be intrinsically out of equilibrium even in the absence of applied forces and external fields \cite{AM2010,AMINTRO2015}. Such features lead to the appearance of  common properties to any active matter system 
%that make it different from any other kind of soft matter
\cite{AM2012} such as the emergence of collective structures with qualitatively different behaviors from individual ones\cite{stark2016,doi:10.1146/annurev-physchem-050317-021237}, out-of-equilibrium phase transitions from disordered  to ordered systems and vice-versa (\textit{coarsening}\cite{stenhammar} and \textit{clustering}\cite{example,CollMot_ABP_Attr_2D,redner2013,gonnella2014}), pattern formation at the mesoscale\cite{Cates11715,Bartolo2019,solon2015}, special mechanical and rheological properties\cite{Martinez17771,Portela2019,gonnella2019}, novel fluctuation statistics with respect to equilibrium and other non-equilibrium systems\cite{Woillez2019,nemoto2019,solon2018,fodor2016}. \\

\noindent To understand, predict and even reproduce these active matter features, the most common procedure is to take a bottom-up approach and, from the individual behavior of the constitutive parts, describe  the collective properties of a system by  means of statistical physics tools. Individual behaviors, however, are in general quite complicated and a first modelling step is needed, thus boiling down the elements of a system to  moving particles obeying  few straightforward rules that replicate the observed dynamics while keeping the number of free parameters fairly low \cite{AM2012}. \\

\noindent Some models  define the particles' dynamics from the speed's direction, establishing aligning rules  leading to collective organized motion \cite{SVM,ColMo}, while others focus on the speed, making particles move slower where the density is higher,  leading to aggregation events, (Motility Induced Phase Separation - MIPS) \cite{MIPS,MIPS2}. An extreme case of  MIPS in non-aligning active systems is Excluded Volume (EV), which enforces that two different particles cannot overlap, also leading to aggregation events when particles are locally trapped together. A  simple implementation of EV is the Persistent Exclusion Process (PEP) introduced in Ref.\cite{RODRI1}. This  consists of a discrete on-lattice model of self-propelled particles obeying run-and-tumble dynamics with the only added condition  that two different particles cannot occupy the same lattice site at the same time. 
%It thus constitutes a simple model where 
Thus, the whole dynamics can be expressed only in terms of two parameters:  the tumbling rate $\alpha$ and  the number density $\phi$. Interestingly, this is enough  to detect clustering and ordering in the system. This model has later been expanded to a finite maximum occupation number per site \cite{RODRI2},  leading to the existence of three different phases (gas, \textit{clusters} and solid). Recently, the PEP model has also been used to explain wetting transitions in bacterial populations \cite{RODRI3}. \\

\noindent Following a similar idea, there exists a number of studies on pattern emergence and self-organization in active systems resulting from the competition between self-propulsion and excluded volume interactions reporting the effects of alignment interactions \cite{CollMot_ABP_Polar} or attractive forces \cite{CollMot_ABP_Attr_1D,CollMot_ABP_Attr_2D,Matas2015,Sarkar2021} both in 1D and 2D. In particular, Ref.\cite{CollMot_ABP_Attr_1D} gives a very interesting example where simple attraction rules and excluded volume lead to complex aggregation dynamics even in one dimension. Indeed, not only does the model presented in Ref.\cite{CollMot_ABP_Attr_1D} allow for aggregation, but, more importantly, leads to the emergence of 
%moving flocks of particles (i.e.
 motile clusters. 
 It is thus one of the few existing modelling attempts to describe collective migration and swarming clusters without alignment interactions. \\

\noindent Collective particles' motion is indeed a key element of active matter, often observed in natural phenomena and of which there exist plenty of experimental studies \cite{SWARMING,CollMigr}. Most often, such dynamics are replicated via alignment interactions between  particles as in the Vicsek model \cite{SVM}. There is evidence, however, that  in  a one-dimensional system of self-driven particles, such alignment  is not needed   to display collective aggregated motion, also  achieved  with attraction forces \cite{CollMot_ABP_Attr_1D}. \\

\noindent In the same spirit, it has been recently shown that self-propelled droplets, confined in a one dimensional micro-fluidic channel, experience a collective dynamics characterised by flocks of neighbouring clusters. This phenomenon is the result  of the interplay between velocity fluctuations and the absence of Galilean invariance \cite{illien}. Cluster condensation takes place as a transient phenomena which  slows down the dynamics, before the system settles into a homogeneous aligned phase. \\

\noindent Other works have recently also turned to the dynamics of one-dimensional systems of active particles (both the works by Dolai \textit{et al.} \cite{indios} and Caprini \textit{et al.} \cite{caprini} are relevant examples) revealing universal scaling and alignment behaviours in one-dimensional active systems. Whether active systems in discrete and continuous space belong to the same universality class is a matter of debate \cite{dittrich2020}. Note that Active Brownian Particles cannot be rigorously defined in 1D since the direction of motion cannot change continuously 
\cite{Cates_2013,MIPS}. \\

\noindent In this work we study a one dimensional system of active run and tumble particles, both on- and off-lattice, for repulsive (excluded volume) and attractive (Lennard-Jones) interactions. We also implement additional dynamics that enables explicit cluster movement in the on-lattice case (in the off-lattice case cluster motility arises naturally from the individual particle dynamics). To characterize the phase behaviour of these systems we look at different analysis tools such as the cluster size distribution (CSD), the fraction of jammed particles (J) and the mean cluster size (M).

\section{\label{Model}One Dimensional Models}
\noindent In this work we consider a one-dimensional system of self-propelled particles obeying run-and-tumble dynamics both on- and off-lattice, considering repulsive (excluded volume) and attractive (Lennard-Jones like) interactions \textcolor{black}{for two reference densities: $\phi=0.8$ and $\phi=0.1$, to see the qualitative difference between an active liquid and and an active gas, where interactions are scarcer.} Under such dynamics, particles can aggregate in clusters. Where the details of the model only allow static clusters to arise naturally, we explicitly introduce a cluster dynamics where clusters are not at rest but can move according to the individual dynamics of their constituent particles.

\subsection{\label{sec:ON}On-lattice system}

\subsubsection{\label{sec:ONpartDyn}Particle dynamics}
\begin{figure}[h!]
\centering
\includegraphics[width=\columnwidth]{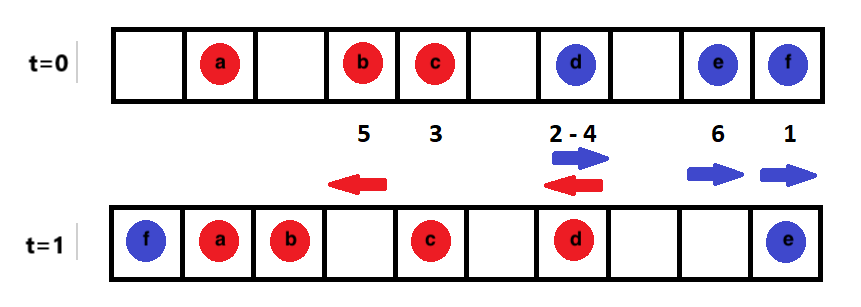}
\caption{Visual representation of the system of $L=10$ and $N=6$ at two consecutive time steps ($t=0$, top row and $t=1$, bottom row). Left-pointing particles are printed in red and right-pointing particles in blue. The numbers indicate the order of updates and the arrows the direction of motion while particle ids are labeled a-f.}
\label{fig:partDyn}
\end{figure}

\noindent We consider a one-dimensional realization of the Persistent Exclusion Process (PEP) model reported in Ref. \citenum{RODRI1} for $10^8$ simulation steps. The system consists of \textit{N} particles on a lattice of \textit{L} sites, whose density is $\phi$ ($N=\phi L$). The size of the lattice sites is the same as that of the particles, $\sigma$, which we fix equal to 1. Particles only move on the lattice sites and are characterized by their position $x_i$ and swimming direction $d_i = \pm 1$:   $d_i = +1$ if a particle $i$ moves to the right (in blue  in Fig. \ref{fig:partDyn}) and $d_i = -1$ if a particle $i$ moves to the left (in red in  Fig. \ref{fig:partDyn}).

\noindent Particles move according to the following scheme mimicking run and tumble dynamics, where both particles positions and directions of motion are updated over discrete time steps. Periodic Boundary Conditions (PBC) are enforced on the system.

\begin{enumerate}
    \item At each time step $N$ particles are chosen at random, allowing for repetitions, and are updated sequentially.
    \item Each of these has a swimming direction (particle orientation) $d_i = \pm 1$ which can be redefined with probability $P_{\text{tumble}}=\alpha$ (the tumbling probability) every time the particle is updated.
    \item Next, the interaction force is computed only if the attractive interactions are included, as discussed below. If only repulsive interactions are present this step is skipped.
    \item We thus define a new moving direction $d_{\text{jump}} = \pm 1$ for each particle as the sign of the resulting total force it experiences (propulsion plus attractive, when present) and a jumping probability\footnote{Note that in general $P_{\text{jump}}\neq P_{\text{tumble}}$ since in some cases the swimming direction of the particle may be pointing in the opposite direction of its movement.} $P_{\text{jump}}$ as the magnitude of such force normalized by the maximum possible force $F_{\text{max}}$. The specific definitions of these forces for each system are discussed in below.
    \item Each particle then performs a one-site jump in the new moving direction $d_{\text{jump}}$ with probability $P_{\text{jump}}$ only if the landing site is empty (excluded volume is enforced). Otherwise, the particle stays in its original position.
\end{enumerate}

\noindent In Fig. \ref{fig:partDyn}, we present an illustrative example of such a system of purely repulsive particles at two consecutive time steps: $N=6$  particles are updated sequentially (as indicated). To give a few examples:  1) one particle \textcolor{black}{(labeled \textit{d}) is updated} twice, while another \textcolor{black}{(labeled \textit{a}) is not.}; 2) particles pointing to an occupied site do not move \textcolor{black}{(see particle labeled \textit{c})};  3) a particle updated twice ends up on its initial site, since it has changed direction (a tumble occurred) during its second update \textcolor{black}{(see particle labeled \textit{d})}, before the jump was performed. 

\subsubsection{\label{sec:ONpot}Short-range attractive and repulsive potentials}
\begin{figure}[h!]
\centering
\includegraphics[width=\columnwidth]{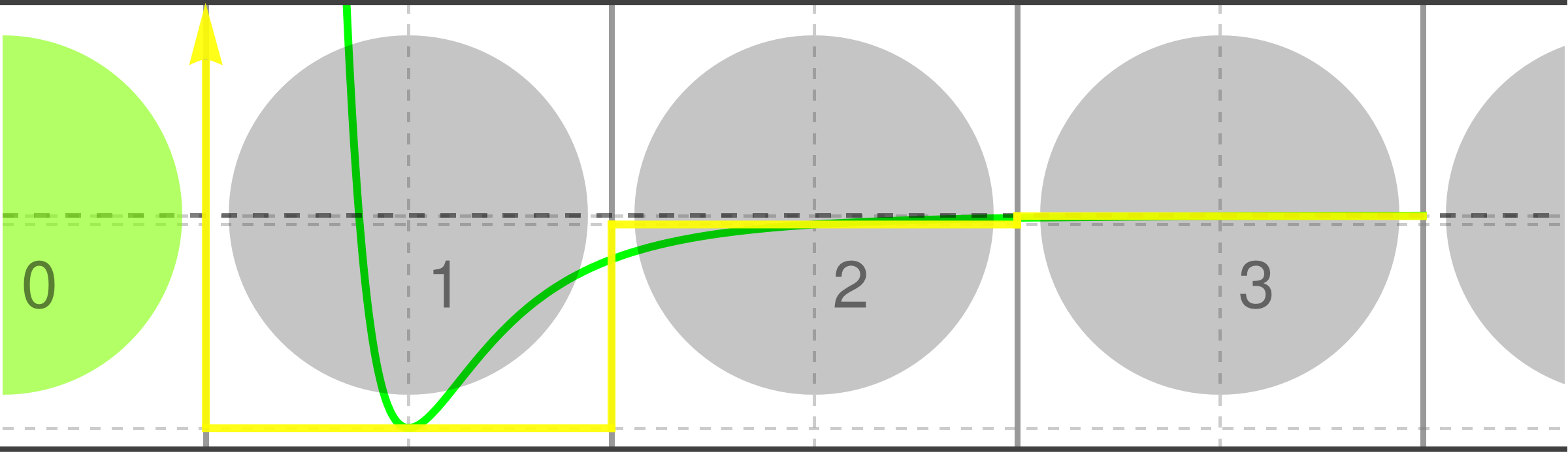}
\caption{Lennard-Jones potential at particle $0$ (in green): the value chosen for each lattice site ($i=1,\,2,\,3\dots,\,r_{\mathrm{cut}}$) inside the influence of the potential, is just the value of the potential at that point: $V(1,\,2,\,3)=-\varepsilon,\,-0.0415\,\varepsilon,\, -0.0025\,\varepsilon$ respectively.}
\label{fig:partPot}
\end{figure}

\noindent To model short-range attractive interactions between particles we define a modified Lennard-Jones potential (truncated, shifted and offset - TSOLJ - $V_{TSOLJ}$) vanishing at a cutoff distance of $r_{\text{cut}} = 3\sigma$ to capture second neighbour interactions in this discrete system. The potential is also shifted in the $r$ coordinate in order for the minimum to be in $r_{\text{min}} = \sigma$ instead of $r_{\text{min}} = \sqrt[6]{2}\sigma$ as the usual Lennard-Jones potential ($V_{LJ}$): $r_{\text{off}}(r)=r+\text{sgn}(r)(\sqrt[6]{2}-1)$, see Fig. \ref{fig:partPot}. Attractive forces are thus computed by summing the contributions of neighbour particles with $F(r) = -\partial_r V_{TSOLJ}(r)$. Repulsive interactions, on the other hand, are implemented via the excluded volume algorithm described in point 5 of the previous section. 

\noindent The TSOLJ potential introduces a new variable: $\epsilon$, the depth of the potential well. Therefore we expect the steady-state of the system to be completely determined by $\alpha$, $\phi$ and $\epsilon$. Constructing phase diagrams of the system will thus allow to better understand the interplay between these three elements.

\subsubsection{\label{sec:ONclustDyn}Cluster dynamics}

\begin{figure}[h!]
\centering
\includegraphics[width=0.75\columnwidth]{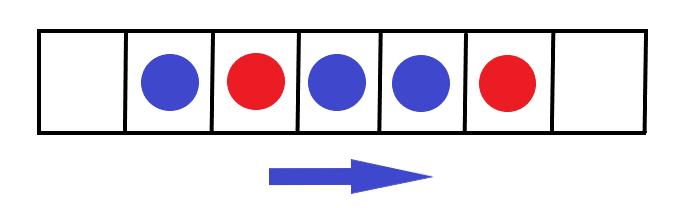}
\caption{Visual representation of a five-particle cluster and its resulting jumping direction. Left-pointing particles are printed in red and right-pointing particles in blue. The arrow indicates the resulting cluster jumping direction. }
\label{fig:clustrDyn}
\end{figure}

\noindent To capture collective dynamics, we propose to modify the classic PEP model \cite{RODRI1} by introducing a cluster motion, independent of the individual dynamics. To start with, we identify a cluster of particles as a group of two or more neighbouring particles "trapped" in a certain position (with particles at the  boundaries having opposing directions, as in Fig. \ref{fig:clustrDyn}).

\noindent To implement cluster mobility, at each time step we carry out the following steps:
\begin{enumerate}
    \item We evaluate whether particles belong to a cluster.
    \item We  compute the cluster's direction of motion $D_C$. by a majority rule, summing each  particle's direction  $D_C = \frac{\sum d_i}{|\sum d_i|}$ ($d_i \pm 1$).
    \item Each cluster will move  with a given  probability $P_C = \frac{|\sum d_i|}{l}$ (where $l$ is the size of the cluster), independently of the rest of the dynamics, and only once all  particles positions have been updated.
\end{enumerate}

\subsection{Off-lattice system}

\noindent As opposed to the previous model, we also study a continuum system. In this scenario we consider a one dimensional system of active  particles described  by their position $x_i \in [0,L]$ (a continuous variable) and their propulsion direction $d_i = \pm 1$. As in the on-lattice model, $i \in [1,N]$ denotes the particle and $L$ the size of the system and PBC are  implemented. For this continuous system we implement two different dynamics, aiming to unveil universal features of the one dimensional system. The first is a classical implementation of Langevin dynamics (underdamped) with elastic collisions, for which we make use of the molecular dynamics software LAMMPS \cite{lammps}. The second is a typical Brownian dynamics (overdamped) of particles propelled with a force $F^p$ \cite{Romanczuk_2012} for which we develop our own code in $C$. The Langevin dynamics is described in the main text, whereas the Brownian dynamics will be left for discussion in the supplementary information.

\subsubsection{Particle dynamics}
\noindent First we consider particles subject to Langevin dynamics. The force that each particle feels,

\begin{equation}\label{langevin_dynamics}
     \frac{m}{\gamma} ~\ddot{x} = -\dot{x} + \frac{D}{k_B T} (F^p+F^i) + \sqrt{2D} \boldsymbol{\eta} 
\end{equation}

\begin{equation}\label{langevin_dynamics}
     m\ddot{\vec{x}}\, =\, -\,\gamma\dot{\vec{x}} \,+\, F_p\hat{\theta} \,-\, \nabla U \,+\, \gamma\sqrt{2D}\, \vec{\boldsymbol{\eta}}(t)
\end{equation}

\noindent where $F^p$ is the self-propelling force \textcolor{black}{of constant magnitude $|F_p|=1$, which randomly changes its direction after a typical reorientation time $\tau_r = 1/\alpha$}, and $F^i$ are the conservative forces at play. $\boldsymbol{\eta}$ is a delta-correlated white noise. In this equations the friction coefficient is defined as  $\gamma = k_B T / D$. We fix $m=\gamma =\sigma = 1$, where $\sigma$ is the diameter of the particle, and $D=k_B T=10^{-4}$. We choose this low temperature value to allow comparison with the on-lattice system but not a lower one to avoid losing the ergodicity of the system when strong attractive interactions are present.\\

\noindent The dynamics is actually controlled only by two dimensionless parameters:
\begin{equation}
    Pe = \frac{v \tau_r}{\sigma}\qquad\text{and}\qquad \xi = \frac{F^p\sigma}{\epsilon}.
\end{equation}
The first is the Péclet number \cite{Jose_Peclet_def}, $Pe$, which captures the contributions of activity relative to diffusion (here, $v = F^p D/k_B T$ is the self-propelling velocity). The second, $\xi$, is the strength of the interaction potential relative to the propulsion force \cite{Francisco_Xi_Definition}, which defines the ratio between the active and conservative forces, it is defined only for finite $\epsilon$, being $\epsilon$ the amplitude of the potential.\\

\noindent\textcolor{black}{It should be noted that care needs to be taken when choosing how to vary the Péclet number, since varying different parameters (for instance $v$ or $\tau_r$) leads to different state diagrams and/or dynamical properties\cite{Jose_Peclet_def,caprini-varyingPe}. Throughout this work we choose to vary the Péclet number by changing the reorientation time, or equivalently, the tumbling rate $\alpha=1/\tau_r$.}

\subsubsection{Short range attractive and repulsive potentials}

\noindent We will first consider only excluded volume interactions, for which we set the interaction to a WCA \cite{wca} potential, which is a truncated and shifted Lennard-Jones potential with the cutoff chosen at the minimum of the potential, $r_{\text{cut}}=\sqrt[6]{2}\,\sigma$. Later on we will also consider attractive interactions, for which we use another modified Lennard-Jones potential (Truncated and Shifted - TSLJ, with $r_{\text{cut}}=2.5\,\sigma$), analog to the one used in the on-lattice system \footnote{Note that this implementation of a Lennard-Jones potential (TSLJ) is different from the one for the discrete model (TSOLJ) in that the cutoff is smaller here and the minimum is not offset to an integer value. However, both definitions retain the general properties of a Lennard Jones potential, tailored to the particular system where they are implemented.}.
\begin{align}\label{TSLJ_continuous}
    V_{\text{LJ}}(r) &= 
    4\epsilon \left[\left(\frac{\sigma}{r}\right)^{12}-\left(\frac{\sigma}{r}\right)^6\right]\\
    V_{\text{TSLJ}}(r) &=
    \begin{cases}
        V_{\text{LJ}}(r)-
    V_{\text{LJ}}(r_{\text{cut}}),\, &r\leq r_{\text{cut}}\\
    0,\, &r>r_{\text{cut}}
    \end{cases}\nonumber
\end{align}

\subsubsection{Cluster dynamics}
\noindent Because in the present case mobile clusters emerge naturally, we find that explicitly implementing cluster dynamics like we do for the discrete system has little effect on the measured observables. Consequently, after proper verification \textcolor{black}{data not shown}, we have decided not to include an explicit cluster move in the off lattice dynamics for the remaining of this work.

\subsection{Computational Details}
\noindent For all systems and models considered we time average the measured observables in the stationary regime. To discard finite size effects, we have carefully performed simulations for systems with varying number of particles $N \in [250,1000]$ (see Supporting Information) and concluded that no finite size effects were present. Consequently, for the remaining of this work we will only consider systems of size $N=500$ particles, adapting the lattice size $L$ to match the particle density $\phi$. The simulation time, depending on the kind of system, is set in the range $T_{sim} = 10^6 ~- ~10^8$ time steps, with $dt$ in the range of $10^{-3}~-~5\times 10^{-2}$. We simulated the one dimensional off-lattice system using LAMMPs \cite{lammps} for $5\times 10^7$ steps with a time step of $0.02$ and for $500$ particles.\\

\noindent The simulations were carried out using GNU-parallel \cite{gnuparallel} on desktop workstations and using Brigit HPC server \cite{brigit}, the codes are available at our GitHub\cite{github}.

% \noindent Throughout this work we fix our units by setting $m=\gamma=\sigma=1$ and define everything in terms of the particle size $\sigma$, the reduced length unit for all systems. \\

\subsection{Differences between the on-lattice and off-lattice models}

\noindent Even though it would be possible to include the temperature effects in the on-lattice model, we choose to simulate the system at T = 0 to focus on the athermal aspects of the dynamics. In the continuous dynamics, on the other hand, we cannot remove temperature effects completely. Consequently, we set T to a very low value in order to favour the athermal dynamics here as well and work with comparable discrete and continuous systems. We find that setting $D = k_BT = 10^{-4}$ is sufficient to have activity dominated dynamics (high Péclet number) while retaining proper dynamics (avoiding non-ergodic states for instance). \\

\noindent As a consequence, in the on-lattice system, when the attraction strength $\epsilon$ increases above a certain value,  particles  aggregate and the aggregate never breaks.
Whereas in the off-lattice system, the presence of the white noise, although weak, allows for the possibility of particles not only to merge into an aggregate but also to leave  an aggregate (theoretically for an arbitrarily high $\epsilon$). \\

\noindent As it will be explained in the Results section (Section \ref{Results}), above some attraction strength, the two models show different behaviours: the on-lattice system ceases to be ergodic and enters an  absorbing state, whereas the off-lattice system displays slower dynamics when increasing the attraction strength but does not fall into an absorbing state.

\section{Analysis tools}

\noindent Throughout this work we are particularly interested in characterizing the state of aggregation of the system. To this end, we have used the following analysis tools: 1) the cluster size distribution; 2) the fraction of jammed particles, $J$; and 3) the normalized average cluster size $M$. \\

\noindent To measure the cluster size distribution at time step $t$ ($P_{l,t}$) we compute the number of clusters of size $l$ at time $t$ ($n_{l,t}$) and normalize by the total number of clusters at that time.
\begin{equation}\label{size_distr}
    P_{l,t} = \frac{n_{l,t}}{\sum_l n_{l,t}}
\end{equation}

\noindent As  expected, the system relaxes to an out-of-equilibrium steady-state with a constant cluster size distribution $P(l) = \langle P_{l,t}\rangle _t$, as in the  PEP model. Therefore, one needs to wait for the system to reach a steady state in order to be able to measure aggregation. \\

\noindent \textcolor{black}{Defining a cluster in the discrete systems is straightforward: any grouping of two or more particles in direct contact with each other (adjacent lattice sites). However, for continuous systems a cutoff distance between particles needs to be defined in order to determine that they are part of the same cluster. We set such cutoff at $x_{cluster}=1.3 \sigma$ so that we reach further than the potential minimum ($\sqrt[6]{2}\,\sigma$) but not far enough for spurious coarsening effects to arise, as detected above $x_{cluster}=1.4 \sigma$.} \\

\noindent While the cluster size distribution constitutes a key element in the study of the system, it doesn't provide a quantitative measure of the state of aggregation. This is the reason why we introduce the aggregation parameter $J$ corresponding to the average fraction of jammed particles (trapped in clusters) in the steady state. 
Its physical interpretation is straightforward:  the closer $J$ is to $1$,  the more aggregated the system will be. If $J=1$ all particles are trapped in clusters and there are no free active particles in the dilute regions. On the contrary, if $J=0$ all particles are free and no aggregation  occurs. \\

\noindent However, $J$ by itself isn't sufficient to assess whether the system is in a coarsened or clustered state as $J$ can be close to $1$ independently of the number of clusters. Indeed, ideally, all particles could be trapped in a few small clusters  (so the system would be in a clustered state) or in a single large cluster (so the system would be in a coarsened state) and $J$ would still be equal to $1$ in both cases. \\

\noindent Therefore, to properly determine what state of aggregation the system is in, we define another aggregation parameter, $M$, as the normalized average cluster size in  steady state ranging from $0$ to $1$.
\begin{equation}
    M = \frac{1}{N_C}\sum_l P(l) l,
\end{equation}
\noindent Where $N_C$ is the total number of clusters. Again, its physical meaning is very straightforward as the closer $M$ is to $1$ the more coarsened the system will be. \\

\noindent Since both $J$ and $M$ allow to determine the state of aggregation of the system quantitatively, they can be interpreted as  "order parameters" for the aggregation of the system. However, these are strictly not order parameters since we are not dealing with a thermodynamic phase transition, so we will refer to them as aggregation parameters. \\

\section{\label{Results}Results}
\noindent For each system studied, on-lattice, on-lattice with explicit cluster moves and off-lattice, we start with a qualitative description of the structural features and then turn to a quantitative description using the above mentioned analysis tools. 

\subsection{On-lattice}

\begin{figure}[h]
    \centering
    \includegraphics[width=\columnwidth]{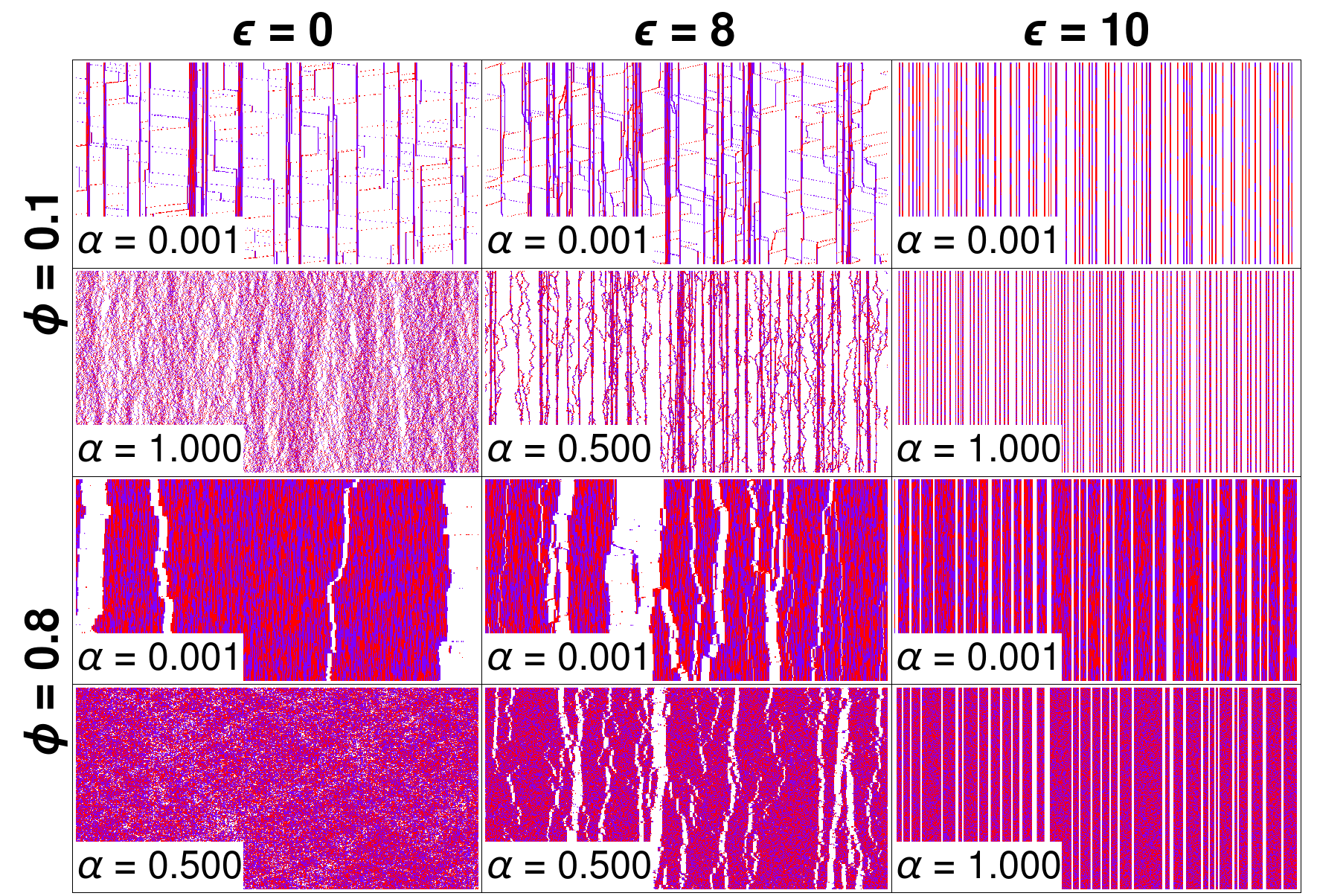}
    \caption{Particle trajectories for the on-lattice system in  stationary state for high (0.8) and low (0.1) densities and three values of the potential intensity $\epsilon$: 0 (repulsive), 8 (attractive  force $\sim$ propulsion force) and 10 (attractive force $>$ propulsion force). High/low values for the tumbling rate $\alpha$ are shown (some of the high values are different for  a better visualisation). In each panel, time flows from bottom to top, each row corresponds to a snapshot of the system at a given time. Left and right moving particles are colored blue and red, respectively.}
    \label{fig:snapsON0}
\end{figure}

\noindent \textbf{Qualitative behaviour.} In Fig. \ref{fig:snapsON0} we represent some particle trajectories of the on-lattice system without explicit cluster dynamics for a wide set of parameters:  the density, $\phi$, the tumbling rate, $\alpha$, and the potential depth, $\epsilon$.

\noindent When particles are repulsive (WCA) the system only forms immobile clusters that never merge (the system is not phase separated). As expected, the system is more aggregated (larger clusters) for high densities ($\phi = 0.8$) and low tumbling rates ($\alpha = 10^{-3}$). Whereas for low densities ($\phi = 0.1$) small clusters form and break continuously, effect that is enhanced for higher tumbling rates ($\alpha = 1$). \\

\noindent When we switch on the attractive interaction and its order of magnitude is comparable to that of the propulsion force ($\epsilon\approx 8$), we find that the aggregation decreases for low $\alpha$ (see for instance $\alpha = 10^{-3}$) with respect to the repulsive case.  For higher tumbling rates, however, (see for instance $\alpha = 0.5$) we find that the presence of attraction increases the overall aggregation of the system. \textcolor{black}{This feature is similar to the reentrant behaviour explained in Ref. \citenum{redner2013} and depicts the counter-intuitive consequence of decreasing the aggregation while increasing the attraction between particles. A detailed explanation of this behaviour is given in the following cluster size distribution section.} We also observe that attractive interactions combined with propulsion provide some mobility to small clusters. \\

\noindent For the strongest attractions studied ($\epsilon=10$), the system enters an absorbing state, in which nothing moves because the attractive force between particles overtakes the propulsion force, and particles that stick together form clusters that never break apart due to the absence of thermal noise. \\

\noindent We can estimate the value of $\epsilon$ for which the first neighbor cannot escape anymore is that in which the attractive force cancels the propulsive force $F_p=1$:
\begin{equation}
    |F_c|=V'_{\text{TSOLJ}}=|F_p|=1\quad\rightarrow\quad \epsilon_{\text{trap}}\approx 8.27,
\end{equation}

\noindent This value is consistent with the one observed in the simulations, where $8<\epsilon_{\text{trap}}\le 8.5$. Indeed, when we consider points in the parameter space beyond this value we find that the system ceases to be ergodic (right column of Fig. \ref{fig:snapsON0}). Consequently, the time average we perform does not correspond to the ensemble average, so the results in this case have to be taken with care.\\

\noindent \textbf{Cluster size distribution.} To better understand the aggregation states presented above let us now turn to the distribution of cluster sizes in the stationary regime. 
\begin{figure}[h!]
    \centering
    \includegraphics[width=\columnwidth]{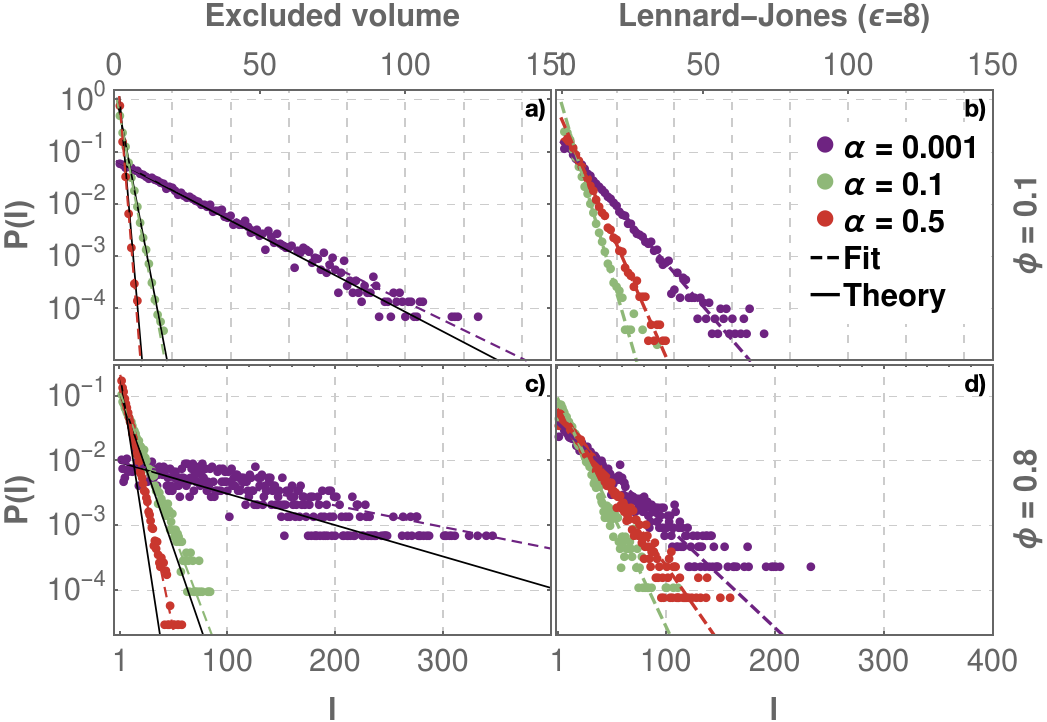}
    \caption{Cluster size distribution $P(l)$ for the on-lattice dynamics where particles interact repulsively (\textcolor{black}{panels a) and c)}) and attractively (\textcolor{black}{panels b) and d)}). Top panels (\textcolor{black}{a-b}) correspond to low density,  bottom panels (\textcolor{black}{c-d}) to high density (note that x-axis scale is different for better visualization). Purple points are data for $\alpha = 0.001$,  green for $\alpha = 0.1$ and red for $\alpha = 0.5$. The solid lines of the left column represent the theoretical prediction of \cite{RODRI1}, Eq. \ref{eq:CSDteo}, and the dashed lines an exponential fit of the data.}
    \label{fig:csdON0}
\end{figure}

\noindent In Fig. \ref{fig:csdON0} we represent the cluster size distributions $P(l)$ for the on-lattice system without any explicit cluster dynamics, where particles are either repulsive (\textcolor{black}{panels a) and c)}) or attractive (\textcolor{black}{panels b) and d)}), where attraction competes with propulsion. They all follow an exponential law. For repulsive interactions ($\epsilon = 0$, panels \textcolor{black}{a) and c)}) we observe the expected behaviour: clusters are bigger for lower tumbling rates and higher densities (purple data, bottom row).
When comparing to Ref. \citenum{RODRI1}, our results for the low density repulsive system (Fig. \ref{fig:csdON0}, top left panel) agree quite well with their theoretical prediction for the cluster size distribution,
\begin{equation}\label{eq:CSDteo}
    P(l)=A_c e^{-\frac{l}{l_c}},\qquad l_c=\sqrt{\frac{2\phi}{(1-\phi)\alpha}}
\end{equation}
\noindent This serves as a check that the model has been implemented correctly. For higher densities (Fig. \ref{fig:csdON0}, panel \textcolor{black}{c)}) our results differ slightly, especially for higher tumbling rates. This may be due to the fact that the assumption of non-interacting clusters is no longer valid in this regime. \\

\noindent Surprisingly, when we switch on the attractive interactions (see \textcolor{black}{\textcolor{black}{panels b) and d)}} in Fig. \ref{fig:csdON0}) the system turns out to be less aggregated for small tumbling rates ($\alpha = 10^{-3}$) and more aggregated for larger tumbling rates ($\alpha = 0.1, 0.5$). Furthermore, now the system for $\alpha=0.5$ is more aggregated than for $\alpha=0.1$. As we can see in panels \textcolor{black}{b) and d)}, the red curve passes over the green one, i.e. now clustering is non-monotonic as a function of the tumbling rate (similar behaviour for a 2D system can be found in Ref. \citenum{Sarkar2021}).\\
 
\noindent This counter intuitive fact, i.e. the system is less aggregated when attraction is present, can be explained by small clusters moving less than in the repulsive case.
As position updating is sequential, cluster motion will only arise from the individual dynamics when cluster's particles are aligned and allowed to temporarily escape it, such that they successively move in their common swimming direction. This second requirement is greatly limited by attractive interactions, which will lower the probability of particles to "swim away" from their cluster, which results in clusters with reduced natural motility.
Such unphysical behaviour, that arises from sequential updating, also motivates the addition of the explicit cluster dynamics described in section \ref{sec:clusterMove}.
Note that this only happens for the lowest tumbling rates (e.g. $\alpha=0.001$) because only in this case is aggregation driven by the mobility of small clusters. Indeed, such mobility depends on the probability of all particles of the cluster to remain aligned in the next step, which corresponds to the joint probability that no particle inside the cluster changes its orientation and decreases monotonically with the tumbling rate $\alpha$ and the cluster size $l$:
\begin{equation}\label{alignmentP}
    P_{\text{align}}=\left(1-\frac{\alpha}{2}\right)^l,
\end{equation}

\begin{figure*}[th!]
    \centering
    \includegraphics[width=\textwidth]{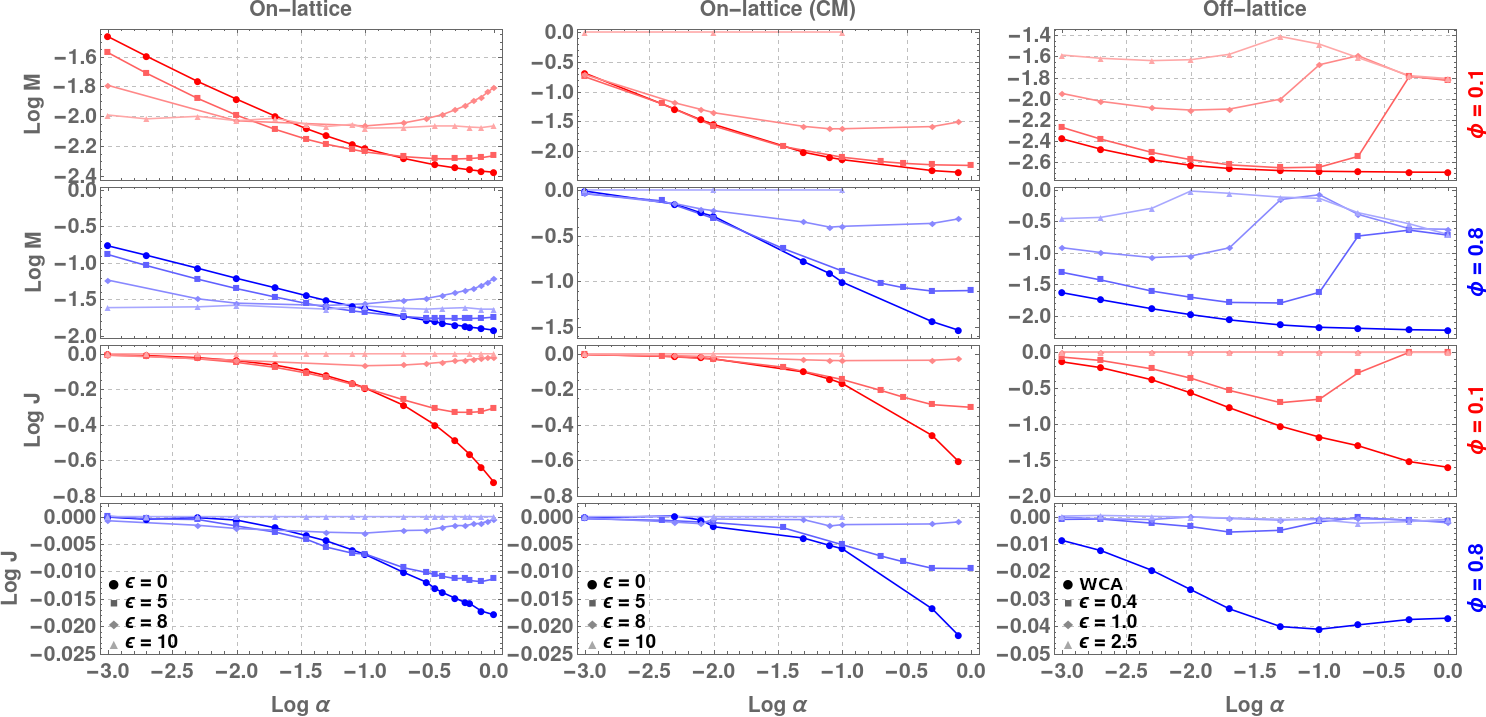}
    \caption{Mean cluster size $M$ (\textbf{top}), and fraction of jammed particles $J$ (\textbf{bottom}) as a function of the tumbling rate $\alpha$ for two different densities ($\phi=0.1$ red, $\phi=0.8$ blue) and four different values of the attractive Lennard-Jones potential depth $\epsilon$ ranging from 0 to 10 in the on-lattice systems and 0 to 2.5 in off-lattice one (for $\epsilon=0$ we consider a purely repulsive WCA potential with $\epsilon=1$). Each column contains one of the three systems studied.}
    \label{fig:JyM_3}
\end{figure*}

\noindent Following this then, for low tumbling rates aggregation is dominated by the mobility of small clusters, which in turn is hindered by the presence of an attractive potential, resulting in lower aggregation for attractive particles than repulsive ones (in this regime of $\alpha$ values). Conversely, for high tumbling rates aggregation is dominated by free particles falling into clusters rather than cluster mobility. This is because in this regime clusters are inherently less likely to remain aligned (and thus collectively move, see Eq. \ref{alignmentP}) and more particles are in the dilute regions because of the lower average time edge particles stay in a cluster (inversely proportional to the tumbling rate). Consequently, in this scenario, introducing attractive interactions favours aggregation as it will increase the cluster lifetime by hindering particle escape. Note that this argument also explains the crossing observed for $M$ (at low $\alpha$ $M$ decreases with increasing $\epsilon$ while at high $\alpha$ $M$ increases with $\epsilon$) in Fig. \ref{fig:JyM_3}, discussed in detail in the following section. \\

{\color{black}
\noindent
\textbf{Parameters} $\boldsymbol{J}$ \textbf{and} $\boldsymbol{M}$. In Fig. \ref{fig:JyM_3} we report results for the mean cluster size and the fraction of jammed particles computed for the on lattice (left-hand column), on lattice with cluster move (center column) and off lattice systems (right-hand column). Particles interact via repulsive or attractive interactions, are characterized by different tumbling rates and suspensions are dense or dilute.\\

\noindent When dealing with the on-lattice system without cluster moves (left-hand column), $M$ decreases as we increase the tumbling rate $\alpha$ between $\alpha = 10^{-3}$ and $\alpha \approx 10^{-1}$ for values of $\epsilon \le 8$ for both $\phi=0.1$ and $\phi=0.8$. This illustrates the fact that as the particles reorient more frequently, it is harder for them to aggregate in larger clusters \textcolor{black}{because they tend to leave the cluster before a new particle comes in increasing the cluster size}. This happens for both densities. At some point between $\alpha=10^{-2}$ and $\alpha=10^{-1}$ the mean cluster size reaches a minimum for $\epsilon=8$ (this corresponds to the overtake of the green curve over the red one for attractive systems in Fig. \ref{fig:csdON0}-right panels). This can be explained since above certain $\alpha$, the probability that a particle that has left a cluster comes back is greater and this is amplified by the presence of the attractive interactions.} \\

{\color{black}
\noindent The crossing observed can be explained \textcolor{black}{by the reentrant behaviour\cite{redner2013} described previously and the fact that} the larger the tumbling rate (lower Péclet number), the more the system behaves as a passive system \cite{filion}, where attraction drives aggregation. However for smaller tumbling rates (larger Péclet number) activity acts as an effective repulsion, thus leading to smaller clusters.} \\

\noindent For $\epsilon > 8$ the attractive force overtakes the propulsion force, so the role of $\alpha$ becomes negligible. Since we are at $T=0$ the system ``freezes'' reaching an absorbing state where particles that stick together never break apart. The system ceases to be ergodic.

\subsection{On-lattice with cluster moves}\label{sec:clusterMove}

\begin{figure}[h!]
    \centering
    \includegraphics[width=\columnwidth]{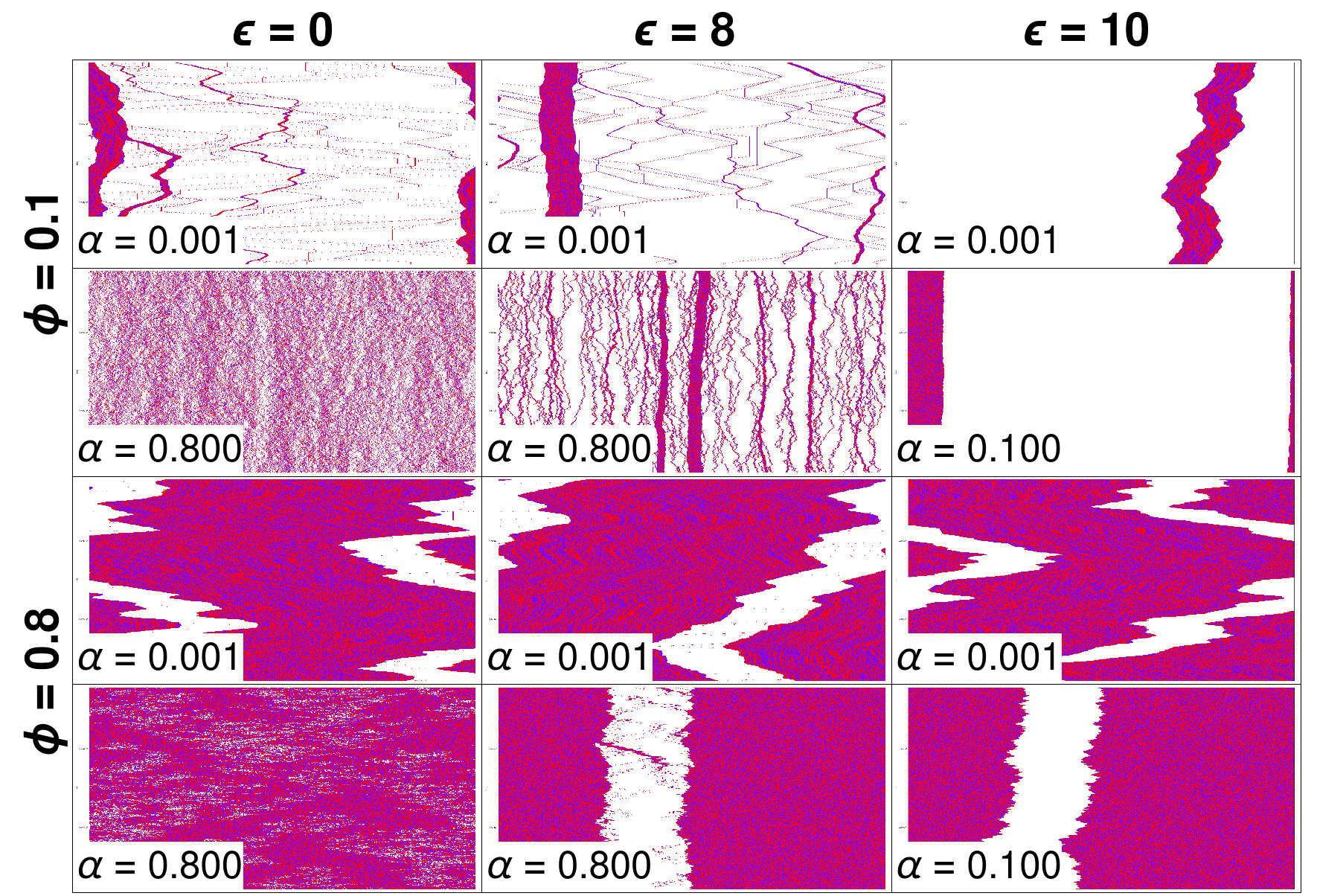}
     \caption{Particle trajectories for the on-lattice dynamics with explicit cluster move in  stationary state for high (0.8) and low (0.1) densities and three values of the potential intensity $\epsilon$: 0 (repulsive), 8 (attractive  force $\sim$ propulsion force) and 10 (attractive force $>$ propulsion force). High/low values for the tumbling rate $\alpha$ are shown (some of the high values are different for  a better visualisation). In each panel, time flows from bottom to top, each row corresponds to a snapshot of the system at a given time. Particles moving left (right) are colored blue (red).}
    \label{fig:snapsON1}
\end{figure}

\noindent
\textbf{Qualitative behaviour.} In Fig. \ref{fig:snapsON1} we represent a few particle trajectories of the on-lattice system with explicit cluster dynamics for a wide set of parameters:  the density, $\phi$, the tumbling rate, $\alpha$, and the potential depth, $\epsilon$.

\noindent For the repulsive system ($\epsilon=0$) and for low density and tumbling rates we now have bigger mobile clusters than in the previous case (compare left panels in Fig. \ref{fig:snapsON1} with left panels in Fig. \ref{fig:snapsON0}), since clusters are now able to move and merge together. However, this effect disappears for higher tumbling rates as activity acts as an effective temperature that prevents cluster formation. For higher densities we find that the system phase separates in two phases: a condensed phase (consisting of one big cluster) that contains the majority of the particles and a dilute phase surrounding the cluster. Again when we increase $\alpha$ the system dissolves completely. Therefore, we suggest that  the cluster motion acts as an effective  attraction. \\

\noindent Now we switch on the attractive potential ($\epsilon = 8$). For low tumbling rates ($\alpha=10^{-3}$) we observe little difference with respect to the repulsive case. This is reasonable because small tumbling rates already act as an effective attraction. For higher tumbling rates ($\alpha = 0.8$), in contrast with the diluted state of the repulsive system ($\epsilon=0$), the low density system  is in a clustered state whereas the  high density system coarsens. Hence, at low density ($\phi=0.1$) attractive interactions trigger a dilute-to-clustered transition, whereas at high densities ($\phi=0.8$) attractive interactions trigger a clustered-to-coarsened transition. \\

\noindent For the strongest attraction ($\epsilon = 10$) analyzed, the absorbing state of the on-lattice case is broken by the mobility of the clusters, we encounter that the system coarsens to form a macroscopic cluster surrounded by vacuum for all four cases: all particles join in one big cluster. \\

{\color{black}
\noindent Therefore, depending on the system parameters $\alpha$ and $\phi$, we  distinguish  three  steady-state regimes:  clustering, coarsening and a transition regime. \\

\noindent Concerning the coarsening regime (see for example 1st and 3rd rows, or 3rd column of Fig. \ref{fig:snapsON1}), the system evolves towards a steady-state of maximum aggregation where all particles get trapped in a single or few very large clusters, surrounded by a very dilute gas of free particles in constant dynamical exchange. \\

\noindent On the contrary, in the clustering regime (see Fig. \ref{fig:snapsON1} for $\alpha=0.8,\,\phi=0.1$ and $\epsilon=0$) the system reaches a state of  minimum  aggregation  where  clusters  dissolve  before merging  together  so  that  particles  are  trapped  in  clusters of very small size surrounded by a dense gas of free particles. \\

\noindent Finally, for a given parameter choice (see Fig. \ref{fig:snapsON1} when $\alpha=0.8,\,\phi=0.1,\,0.8$ and $\epsilon=8$) we find that the system behaves in an intermediate way (transition regime),  with clusters merging together and increasing  in  size  but without reaching their maximum size, either because of their small mobility or because  dissolving too quickly.}
\begin{figure}[h!]
    \centering
    \includegraphics[width=\columnwidth]{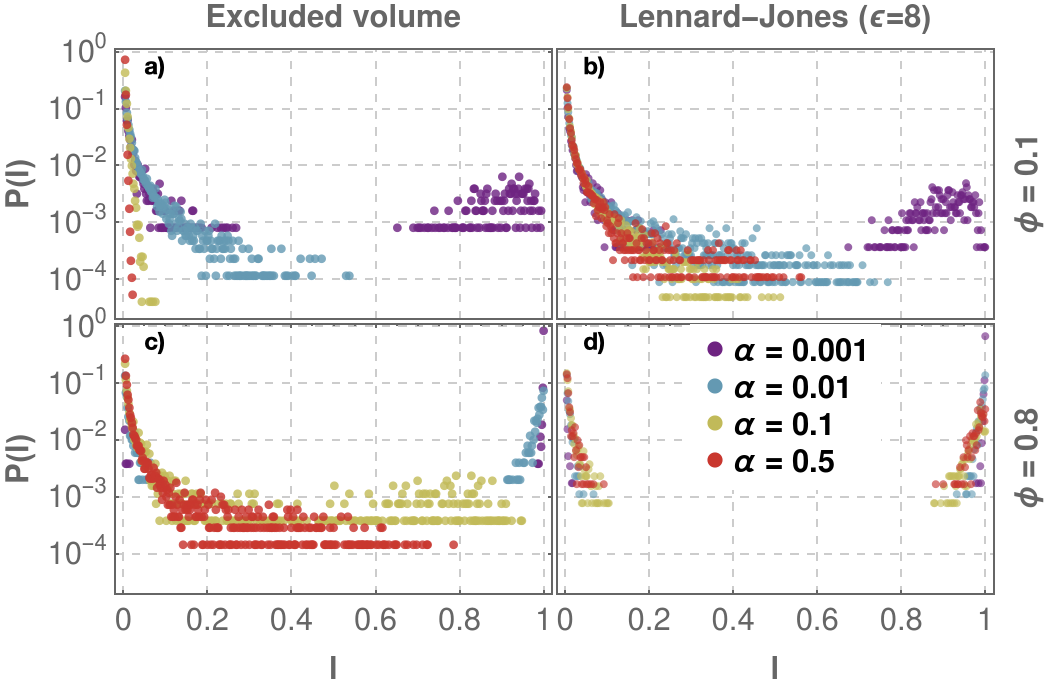}
    \caption{Cluster size distribution $P(l)$ for the on-lattice with cluster move dynamics where particles interact repulsively (\textcolor{black}{panels a) and c)}) and attractively (\textcolor{black}{panels b) and d)}). Top panels (\textcolor{black}{a-b}) correspond to low density,  bottom panels (\textcolor{black}{c-d}) to high density.}
    \label{fig:CSDON1}
\end{figure}\\

{\color{black}
\noindent
\textbf{Cluster size distribution.} In Fig. \ref{fig:CSDON1} we represent the cluster size distributions $P(l)$ for the on-lattice system where an explicit cluster move is implemented. \\

\noindent In the repulsive case (excluded volume), at low density (\textcolor{black}{panel a)}), we identify a clustering regime. This is consistent with what was already detected in the system without explicit cluster dynamics. However, this regime happens only at the highest tumbling rates ($\alpha = 0.1, 0.5$, yellow and red dots). \\

\noindent Note here the single decaying branch in  the cluster size distribution \textcolor{black}{(Fig. \ref{fig:CSDON1}, \textcolor{black}{panel a)})}, which is quite  similar, for this parameter choice, to that observed in the simple PEP model \cite{RODRI1}. \\

\noindent For $\alpha = 0.01$ the distribution is no longer exponential since now particles are sufficiently active to start forming clusters which now can move and merge together creating bigger clusters (entering the transition regime). For even lower tumbling rates, $\alpha=0.001$, we observe a bimodal distribution starting to emerge (coarsening regime). Since now particles spend more time moving in one direction before they tumble, they will form bigger clusters (right peak of the distribution) surrounded by a dilute phase (left peak). Again, these clusters will move, thus absorbing both other clusters and  particles in their way. However, as the cluster size increases, the cluster will move slower. This is the reason why we encounter less clusters of intermediate sizes: they merge forming bigger and bigger clusters until they are so big that they are too slow to keep encountering each other. This leads to the coarsening regime discussed earlier which corresponds to the right peak of the distribution reaching $l/N=1$. \\

\noindent For the attractive and low density case (\textcolor{black}{panel b)}), we observe that interactions drive the system from the clustering regime to the transition regime for $\alpha=0.1,\, 0.5$, increasing the aggregation also for $\alpha=0.01$. This makes sense as one would expect that attraction makes the system more aggregated. Surprisingly, as we also detected in the previous case (sec. \ref{sec:ON}), the system for $\alpha=0.001$ with attractive interactions neither appears to be more nor less aggregated than the repulsive one. This fact suggests that attractive interactions play no role in driving aggregation at %\sout{high}\textcolor{black}{persistent??} 
high activities (low tumbling rates, high persistence length), which is dominated by cluster merging. Also, as for the previous case, the system is slightly more aggregated for $\alpha = 0.5$ than for $\alpha = 0.1$. As explained before, this can be due to the reabsorbing probability of  particles: for low $\alpha$ a particle that evaporates from a cluster may never come back, while for higher $\alpha$ the probability of being reabsorbed into a cluster increases and is amplified by the attractive interactions.\\

\noindent
\textbf{Parameters} $\boldsymbol{J}$ \textbf{and} $\boldsymbol{M}$.
In the central column of Fig. \ref{fig:JyM_3} we represent the mean cluster size, $M$, and the fraction of jammed particles, $J$ for the on-lattice system with cluster move. We observe that the system is overall more aggregated or at least equally aggregated than in the case without explicit cluster move (right column of Fig. \ref{fig:JyM_3}). Even though the functional forms of both $M$ and $J$ curves are similar to the ones obtained without cluster moves. \\

\noindent The effect of adding the cluster dynamics on $J$ does not seem to be as relevant as its effect on $M$, where we  observe that at low tumbling rates the system is much more aggregated than the system without explicit cluster dynamics. Moreover, as we have suggested before, since 1) the cluster move acts as an effective attraction and 2) this effect is especially present for lower tumbling rates, we  also observe that a  higher attraction ($\epsilon=8$) is needed in order to distinguish the $M$ curves among each other, since now aggregation at low $\alpha$ is mostly driven by the explicit cluster dynamics. \\

\noindent As observed in the absence of explicit cluster dynamics, , here we also detect a crossing of the $M$ curves, although it is much more subtle. This can be explained for the same reasons as before: 1) now clusters move explicitly as a whole, so attractive interactions reduce much less the mobility of small clusters (although they do reduce it to some extent since clusters can still move due to their particles being aligned) and 2) the larger the tumbling rate (lower Péclet number), the more the system behaves as a passive system \cite{filion}, where attraction drives aggregation. However, for smaller tumbling rates (larger Péclet number) activity acts as an effective repulsion, thus leading to smaller clusters. Although in this case (where explicit cluster moves are present) the activity-driven aggregation is much stronger since it is amplified by the cluster moves. \\
 
\noindent Moreover, as before, for $\epsilon > 8$ the attractive force overtakes the propulsion force. However, when explicit cluster dynamics is present, the effect of $\alpha$ is more relevant (since it is present in the probability of cluster moves). The absorbing state reached previously is broken due to the explicit cluster dynamics, which leads to coarsening kinetics and a macroscopic cluster, where $M=J=1$ (see  also right column of Fig. \ref{fig:snapsON1}).}\\

\subsection{Off-lattice Langevin dynamics}\label{sec:offlatticeLangDyn}
\noindent In this section we will consider the same system of active particles, subject to a continuum, Langevin dynamics, where inter-particle collisions are not inelastic.
\begin{figure}[h!]
    \centering
    \includegraphics[width=\columnwidth]{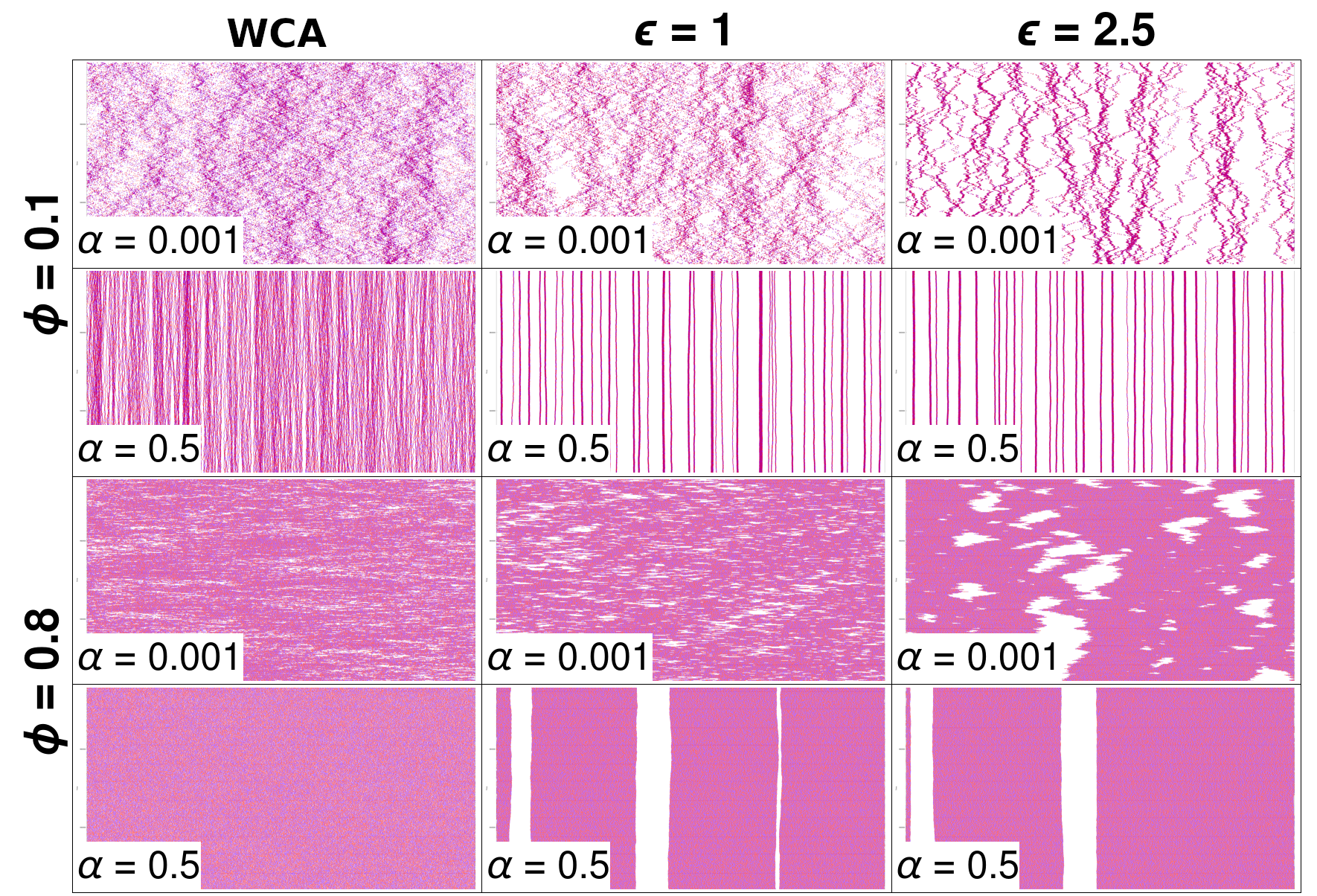}
     \caption{Particle trajectories for the off-lattice system for low (0.1) and high (0.8) densities and three values of the potential intensity $\epsilon$: 0 (repulsive), 1 (attractive  force $\sim$ propulsion force) and 2.5. High/low values for the tumbling rate $\alpha$ are shown (some of the high values are different for  a better visualisation). In each panel, time flows from bottom to top, each row corresponds to a snapshot of the system at a given time. Particles moving left (right) are colored blue (red).}
    \label{fig:snapsOFFlammps}
\end{figure}

\noindent
\textbf{Qualitative behaviour.} In Fig. \ref{fig:snapsOFFlammps} we represent the particles' trajectories for this dynamics for a wide set of parameters:  the density, $\phi$, the tumbling rate, $\alpha$, and the potential depth, $\epsilon$. \\

\noindent When particles are repulsive ($\epsilon = 0$) the system forms clusters that now have some mobility. Since we are in continuous space, particles can ``push'' each other due to momentum conservation during collisions. This is different with respect to what was observed  in the on-lattice model, where one particle at the edge of the cluster could block the entire cluster, even if the rest of the particles were pointing in the same direction against it. \textcolor{black}{This type of behaviour is reminiscent of spatial velocity correlations recently observed in systems of active particles, both in phase-separated
configurations \cite{Caprini_PRL_correlations} or
homogeneous active liquids \cite{Szamel_EPL_correl}.} \\

\noindent As we would expect, the system is more aggregated for high densities ($\phi = 0.8$) and low tumbling rates ($\alpha = 10^{-3}$).
Although under visual inspection the system seems more aggregated for the off-lattice dynamics, the cluster size distribution shows otherwise. Clusters might have longer lifetimes but they are overall smaller than in the on-lattice case. \\

\noindent When we switch on the attractive interaction we observe that the aggregation increases with $\epsilon$ as expected. For low density ($\phi = 0.1$), the system in a cluster state  evolves towards a transition regime (between clustering and coarsening).
For  $\alpha=0.5$ the system dynamics slows down, getting closer to the absorbing state described earlier for the on-lattice system. However, in this case the system does not  completely reach the absorbing state, due to the non-zero temperature and the momentum conservation during collisions. 
For high density ($\phi=0.8$) and high tumbling rate ($\alpha=0.5$), the attraction strength $\epsilon$ leads to a coarsening transition, undetected in the on-lattice system due to the reduced cluster mobility. For $\alpha=0.001$ and $\epsilon=2.5$, the system enters a new phase of small, long-lived and dynamic clusters, undetected in the simple on-lattice case. For off-lattice  attractive dynamics, activity acts as an effective repulsion\cite{PhysRevE.88.012305,example}, reason  why we observe a lower aggregation for $\alpha=0.001$ (high activity)\cite{filion} than for $\alpha=0.5$ (low activity).\\
  
\begin{figure}[h!]
    \centering
    % \textbf{cluster cutoff = 1.5}
    % \includegraphics[width=\columnwidth]{csdOffLammpsfig.png}
    % \textbf{cluster cutoff = 1.3}
    % \includegraphics[width=\columnwidth]{csdOffLammpsfig_Ccut1.3.png}
    \includegraphics[width=\columnwidth]{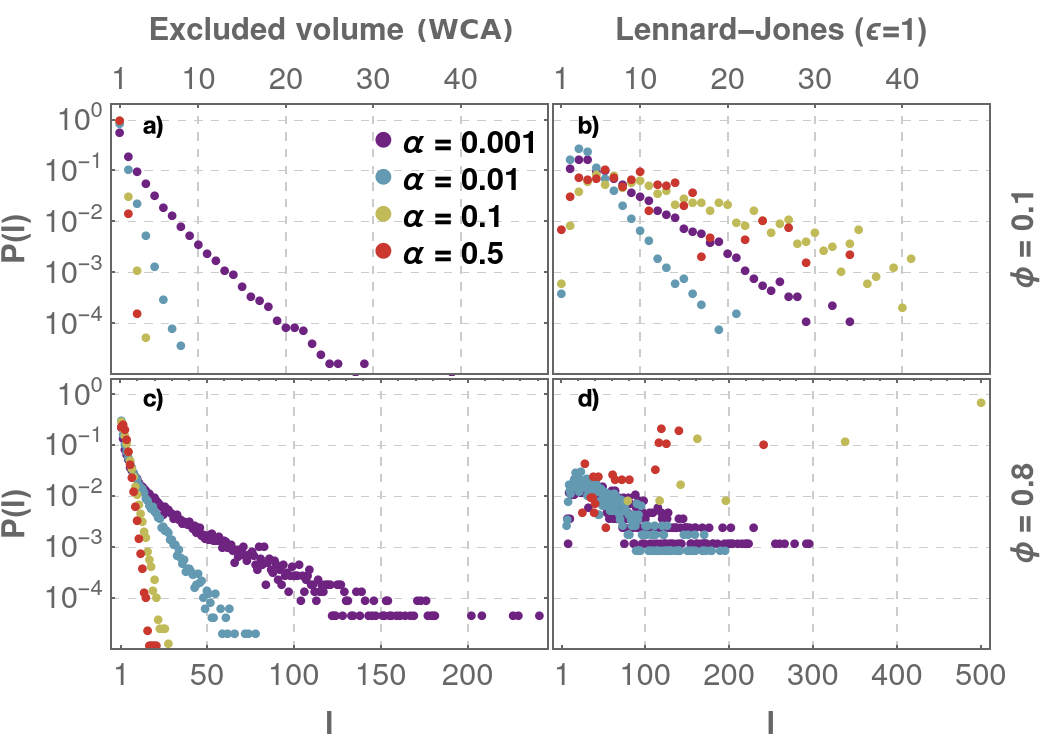}
    \caption{Cluster size distribution $P(l)$ for the off-lattice dynamics where particles interact repulsively (\textcolor{black}{panels a) and c)}) and attractively (\textcolor{black}{panels b) and d)}). Top panels (\textcolor{black}{a-b}) correspond to low density,  bottom panels (\textcolor{black}{c-d}) to high density (note that x-axis scale is different for better visualization).}
    \label{fig:CSDoffLammps}
\end{figure}

\noindent
\textbf{Cluster size distribution.} Fig. \ref{fig:CSDoffLammps} displays the cluster size distribution, $P(l)$, for continuum dynamics  for repulsive (\textcolor{black}{panels a) and c)}) and attractive (\textcolor{black}{panels b) and d)}) interactions at $\epsilon=1$.

\noindent For repulsive  interactions (\textcolor{black}{panels a) and c)} - Fig. \ref{fig:CSDoffLammps}), all curves seem to follow an exponential law, resembling on-lattice dynamics (Fig. \ref{fig:csdON0}), even though in the off-lattice case the slopes are steeper (smaller aggregation overall).
This can be explained since in the off-lattice case  (continuous dynamics)  particles have  more freedom to move than in the on-lattice case (discrete dynamics). \textcolor{black}{For $\phi=0.8$ (\textcolor{black}{panel c)}), the system is obviously more aggregated and starts deviating from an exponential distribution for the lowest tumbling rate, $\alpha=10^{-3}$.} \\%this effect is also reflected in the mean cluster size as we will see later.

\noindent For attractive interactions (\textcolor{black}{panels b) and d)} - Fig. \ref{fig:CSDoffLammps}), the slopes are smaller, corresponding to a higher aggregation, as expected. Note that in the presence of attraction, particle activity acts as an effective repulsion, opposing attraction-induced particle aggregation. This is the reason why the overall aggregation is lower for $\alpha = 0.001$ than $\alpha \geq 0.1$. The specific value of $\alpha$ which separates between  activity-driven and attraction-driven aggregation is yet to be explained. This can also be seen when comparing attractive and repulsive systems at the same tumble rate: the increase in aggregation due to the attraction ($\epsilon = 1$) is less noticeable for higher activity (lower $\alpha$).
For higher density (\textcolor{black}{panel d)}) and high tumble rate (low activity) the system rapidly coarsens into a single or a few clusters, due to the combination of high density, attractive interactions and almost passive dynamics (see the green and red curves in Fig. \ref{fig:CSDoffLammps}). \\

\noindent
\textbf{Parameters} $\boldsymbol{J}$ \textbf{and} $\boldsymbol{M}$. The right column of Fig. \ref{fig:JyM_3} displays the mean cluster size, $M$, and the fraction of jammed particles, $J$ for off-lattice dynamics. The mean cluster size, $M$, increases with epsilon, regardless of $\alpha$, as expected, which is one of the main differences with respect to on-lattice dynamics, where increasing $\epsilon$ decreases $M$ for $\alpha\lesssim 0.1$. \\

\noindent Particle aggregation as a function of $\alpha$ exhibits different trends depending on  the magnitude and type of particle interactions. For repulsive interactions, aggregation decreases with increasing $\alpha$ as in the absence of attraction, high activity (low $\alpha$) drives aggregation. This can be observed both in $J$ and $M$ (Fig. \ref{fig:JyM_3}). Note that both for the on-lattice and for the off-lattice systems aggregation monotonically decreases with the tumbling rate for the repulsive system.
As soon as the attraction is switched on aggregation increases.
However, its dependence on particle activity (tumbling rate) becomes non-monotonic due to the interplay between inter-particle attraction and self-propulsion. For low $\alpha$, the particles have high persistence and the activity dominates over the attraction. The role of activity is twofold: it dampens or screens the aggregation induced by attractive interactions but at the same time it also drives aggregation induced by the persistence of particles motion. Hence the aggregation curves ($J$ and $M$) decay with $\alpha$ in a similar way to that of the repulsive case. As the tumbling rate increases, the persistence of the particles drops, and the attraction begins to take over (more so for higher values of $\epsilon$) until the dynamics are completely dominated by the attractive interactions for high $\alpha$ (similar to a passive system). In this extreme case we observe that the fraction of jammed particles $J$ reaches $1$ for all simulations and the average cluster size $M$ collapses to a single value as a function of $\phi$. Individual particle activity is completely hindered here and aggregation (beyond initial cluster formation) is determined by emergent collective cluster mobility (which decays with cluster size). Consequently, the crossover between these two extreme scenarios (activity or attraction dominated dynamics) is characterised by a minimum in overall aggregation $J$ and $M$. We have to mention that care has to be taken in the region of high $\alpha$ and high $\epsilon$, since the dynamics of clusters slows down strongly and we cannot assure that the steady state is reached. The curve collapse is a finite size effect. Clusters evolve more slowly as their size increases; the times reached in the simulations cannot capture the evolution of clusters beyond a given threshold size. Accordingly, the curve of highest attraction ($\epsilon=2.5$) sets the saturation of the aggregation parameters, which decreases as $\alpha$ increases. This dependence is consistent with the fact that clusters move slower for higher tumbling rates and/or stronger attraction so aggregation events become harder to capture over the simulation times considered. Furthermore, the saturation values depend on the density of the system $\phi$, which relates to the fact that clusters have to travel on average a greater distance before collision in low density scenarios. \\

\noindent The distinction between individual particle dynamics and collective cluster dynamics is also observed on-lattice dynamics. Indeed, when explicit cluster moves are implemented (see middle column in Fig. \ref{fig:JyM_3}) we observe a sharp increase in aggregation with the attraction strength. \\

\noindent In conclusion the differences between on-lattice and off-lattice dynamics cannot be attributed to inertial effects present in the latter. We have shown that a system with overdamped Langevin (Brownian) dynamics, where  inertia is not taken into account, shows similar results to those obtained for the  Langevin system (see the supplementary information). \\

\textcolor{black}{\section{\label{Discussion}Discussion}}

\noindent We have studied a one dimensional system of active run-and-tumble particles whose dynamics is either on-lattice or off-lattice (Langevin-like). To unravel the role played by the inter-particle interactions, we have considered the effect of repulsive and attractive short range interactions. \\

% Commmon to all models
\noindent While important differences between discrete and continuous dynamics, a few common features are of particular importance. As expected intuitively, increasing particle density $\phi$ drives aggregation across all the different dynamics and scenarios explored. The effects of particle activity (self-propulsion) on aggregation are largely independent of the details of the dynamics for non-attractive particles. All repulsive systems display monotonically decaying aggregation for increasing tumbling rate $\alpha$ (inversely proportional to the particles' persistence length and a measure of the activity in the system). However, once attractive interactions are introduced, the steady-state, which is completely determined by the interplay between these and the particles' activity, becomes highly dependent on the details of the dynamics. \\

\noindent In essence, aggregation can occur in two ways, each of which will dominate the system's dynamics under different conditions: 
\begin{enumerate}
    \item \textbf{Particle capture in clusters:} when a free particles encounters an existing cluster and falls 'jammed' inside. For this mechanism we distinguish two regimes: a) For non-attractive particles, persistence ($\sim 1/\alpha$) favours this type of aggregation. b) For attractive particles, the interaction strength favours particle trapping but the activity favours escaping and the trade-off between the two will determine how dominant this process is (beyond the initial cluster nucleation).
    \item \textbf{Custer merging:} when two existing clusters encounter one another. This requires clusters to be mobile and is therefore what depends most on the details of the dynamics. For discrete dynamics without explicit collective cluster motion (what we call 'on-lattice' systems) it is very difficult for these clusters to achieve motility as this requires all particles to be aligned and for the sequential update to occur in the right order. If cluster move is allowed, clusters become very dynamic and this aggregation mechanism can be dominant in the system. Finally, for continuous dynamics, because particles can push one another, mobile clusters are not as rare as in the discrete scenario but still require some degree of alignment and will not display the same dynamics as the 'on-lattice with cluster move'.
\end{enumerate}

\noindent Let us now summarize how the two  aggregation mechanisms act for the different analyzed dynamics.\\

\noindent \textbf{Discrete ``on-lattice'' dynamics.} First, for repulsive interactions we recover the results obtained by Ref. \citenum{RODRI1}. Second, in the presence of attractive interactions, interactions do not break the exponential shape of cluster size distribution but change the global aggregation in different ways depending on the activity regime. This is because, as explained above, in this type of system clusters are mostly immobile, which prevents cluster merging. Consequently, aggregation is dominated by particle capture and: a) for low $\alpha$ (high activity), since in this system attraction decreases cluster mobility, it leads to a decrease in the mean cluster size $M$: activity-driven aggregation. b) for high $\alpha$ (low activity), attraction leads to an increase of $M$ because the system behaves as a passive system controlled by attraction-driven aggregation. When the attractive forces dominate over propulsion, the system is no longer ergodic, due to the absence of thermal fluctuations, and enters an absorbing state. \\

\noindent To improve the cluster mobility with respect to the above scenario, we implemented an explicit cluster move in the on-lattice system. This fundamentally changes the aggregation dynamics of the system, which becomes easily dominated by cluster merging and can achieve a novel coarsened state for a range of parameters (inter-particle interactions or particle density) as long as particles display sufficiently persistent dynamics (low $\alpha$). In this regime (e.g. $\alpha = 0.001$) attractive interactions have almost no effect on the steady state of the system because aggregation is dominated by collective dynamics (cluster motion). However, as we decrease particle activity, the global aggregation (both $J$ and $M$) is reduced. This decay is less pronounced for higher attraction strengths $\epsilon$. This is because as the attraction overtakes the activity and clusters become more stable, merging aggregation mechanisms result in larger clusters in the system and can eventually drive complete particle aggregation in a single stable cluster. In this regime, the cluster size distribution is bimodal, rather than exponential, accounting for the dilute (left peak) and aggregated regions (right peak). Note that when cluster moves are implemented, the system is never trapped in an absorbing state. \\

\noindent As a result, in this system we detect three different regimes depending on the choice of parameters. 1) A clustering regime, with the system consisting of small dynamic clusters. 2) A coarsening regime, with the system consisting of few large particle aggregates, surrounded by a dilute region of particles and small clusters. 3) A transition regime, with the system consisting of a phase in between clustering and coarsening: the clusters life time is sufficiently long for them to start merging into bigger clusters. \\

\noindent In order to understand the relevance of particles being confined to a lattice, we extended the above models to continuous space, where particles evolve over time according to a Langevin dynamics. \\

\noindent \textbf{Continuous ``off-lattice'' system.} Because in the Langevin dynamics clusters can move naturally, cluster merging can further drive aggregation and we observe similar steady state 'phases' to those of the on-lattice model with cluster moves, with both clustering and coarsening. In this case, however, the range of parameters where the system coarsens into a single large aggregated body is much more limited (e.g. we only detect such a regime for large particle density $\phi$). This is because, instead of being enforced, here the collective dynamics required for this level of aggregation only arise under certain conditions where clusters are sufficiently stable (i.e. attractive interactions are sufficiently strong to hinder escape events). \\

\noindent As a result, the aggregation in the steady state is completely determined by the interplay between activity, attraction and thermal noise. Indeed, in the absence of attraction, activity drives aggregation, with both the fraction of jammed particles $J$ and the mean cluster size $M$ being higher for lower tumbling rate $\alpha$: the aggregation dynamics are completely dominated by free particles captured by clusters. However, when attraction is introduced, the effect of activity is inverted as it is now favouring particles escaping clusters, which are in turn stabilised by the interactions. This results in cluster merging taking over for sufficiently low activity or sufficiently high attraction, which drives aggregation back up. \\

\noindent It is worth noting that the greater number of allowed states of the system (due to continuum space) leads to a lower global aggregation than its lattice counterpart in the absence of inter-particle attraction. The robustness of the results described are further confirmed by the study of continuum, overdamped dynamics, which displays analogous behavior in the absence of inertial effects (see SI, section I). \\

\textcolor{black}{\section{\label{Conclusion}Conclusion}}

%Final remarks
\noindent Overall, we find that one-dimensional systems of interacting (attractively or repulsively) run-and-tumble particles display very rich and complex aggregation dynamics resulting in a variety of steady state morphologies. The most common of these is the \textit{clustering} regime, characterised by the formation of many small particle aggregates in constant dynamic exchange with the gas phase surrounding them. When such clusters are sufficiently long-lived and display collective dynamics the system evolves towards a novel \textit{coarsened} phase where all particles aggregate into a single or a few large clusters surrounded by a gas phase of free particles whose density decreases with increasing inter-particle attraction. Steady state aggregation is completely determined by particle density, $\phi$, particle activity, $\alpha$, and interaction, determined by the potential strength, $\epsilon$. The details of the system (i.e. the particular dynamics implementation and the choice between continuous or discrete space) appear to only affect the general results in how they hinder or enable emergent or explicit collective cluster dynamics, which are essential for reaching the \textit{coarsened} steady state described above. \textcolor{black}{Among the studied cases, this state only occurs in two: discrete repulsive with cluster moves at low tumbling rate and continuous attractive at high tumbling rate, since these are the ones that include the two main ingredients for coarsening: long lived and motile clusters. It should be noted that alignment between particles orientations plays no role in the formation of these clusters, although it has been shown that they can promote MIPS\cite{elena-alignment-MIPS}}. This work thus offers a comprehensive view of the dynamics of particles confined to a single dimension, highlighting the diversity and complexity arising from such an \textit{a priori} simple system. \\

\noindent Finally, it is worth mentioning that unlike 2D off-lattice repulsive run-and-tumble systems, in 1D \textcolor{black}{we cannot strictly speak of MIPS at low tumbling rates because, although particles aggregate in clusters with a given size distribution, these do not merge into one dense phase\cite{indios}}. We believe that future work extending the results presented here to two-dimensional systems should prove helpful in untangling the role played by dimensionality in this systems. This research could also be extended by deriving a length scale $l_c$, similar to Eq. (\ref{eq:CSDteo}), that captures the attractive on-lattice clustering behaviour or by considering distributions of the studied parameters instead of fixed values, much in the spirit of Ref. \citenum{Castro2021}.\\

\section*{Conflicts of interest}
There are no conflicts to declare.

\section*{Acknowledgements}
The authors cordially acknowledge useful discussions with Rodrigo Soto and Pablo de Castro.
C. Valeriani acknowledges fundings from MINECO PID2019-105343GB-I00. I. Pagonabarraga acknowledges support from Ministerio de Ciencia, Innovaci\'on y Universidades MCIU/AEI/FEDER for financial support under grant agreement PGC2018-098373-B-100 AEI/FEDER-EU, from Generalitat de Catalunya under project 2017SGR-884, Swiss National Science Foundation Project No. 200021-175719 and the EU Horizon 2020 program through 766972-FET-OPEN NANOPHLOW. R. Brito acknowledges support by the Spanish Ministerio de Economía y Competitividad (Grant number FIS2017-83709-R and PID2020-113455GB-I00).


%\clearpage
%\bibliography{biblio}

\begin{thebibliography}{55}
\bibitem [{bri()}]{brigit}%
  \BibitemOpen
  \href@noop {} {\enquote {\bibinfo {title} {Hpc brigit server at complutense
  university of madrid it services.}}\ }\BibitemShut {NoStop}%
\end{thebibliography}

\end{document}